# Forecasting Managerial Turnover through E-Mail Based Social Network Analysis

Gloor, P. A., Fronzetti Colladon, A., Grippa, F., & Giacomelli, G.

**Forecasting Managerial Turnover through E-Mail based Social Network Analysis**

Gloor, P. A., Fronzetti Colladon, A., Grippa, F., & Giacomelli, G

**Abstract**

In this study we propose a method based on e-mail social network analysis to compare the communication behavior of managers who voluntarily quit their job and managers who decide to stay. Collecting 18 months of e-mail, we analyzed the communication behavior of 866 managers, out of which 111 left a large global service company. We compared differences in communication patterns by computing social network metrics, such as betweenness and closeness centrality, and content analysis indicators, such as emotionality and complexity of the language used. To study the emergence of managers' disengagement, we made a distinction based on the period of e-mail data examined. We observed communications during months 5 and 4 before managers left, and found significant variations in both their network structure and use of language. Results indicate that on average managers who quit had lower closeness centrality and less engaged conversations. In addition, managers who chose to quit tended to shift their communication behavior starting from 5 months before leaving, by increasing their degree and closeness centrality, as well as their oscillations in betweenness centrality and the number of "nudges" they need to send to peers before getting an answer.

**Keywords**

## 1. Introduction

Researchers have been investigating the determinants of employee and managerial turnover for several decades (Holtom, Mitchell, Lee, & Eberly, 2008). Factors such as job satisfaction, economic conditions, and personal motivators are among the variables most frequently reported as leading to voluntary turnover (Egan, Yang, & Bartlett 2004). The literature on turnover recognizes that turnover is not a discrete event, but rather a process of disengagement that can take days, weeks, or months until the decision to leave is made. In this paper, we describe an innovative method to determine who is more likely to leave a company and when. Using a social network approach to collect and analyze data on communication style, we demonstrate the analytical power of traditional social network metrics such as closeness, betweenness and degree centrality (Borgatti, 2005; Wasserman & Faust, 1994), as well as novel indicators such as response time and number of nudges sent and received by employees (Gloor & Giacomelli, 2014).

Looking at the overall 18 months of communication, we focused on patterns emerging during the last five months leading to managers' departure. As suggested by the literature on disengagement at work (Burris, Detert, & Chiaburu, 2008; Kahn, 1990; Luthans & Peterson 2002), employees can be emotionally, cognitively or physically engaged and go through specific phases of active disengagement or alienation. By looking at 5 months prior to voluntary departure, we aimed at capturing the emergence of a communication behavior that would signal the "*decoupling of the self from the work role and people withdrawing and defending themselves during role performances*", which represents the definition of disengagement according to Kahn (1990, p. 694).

The decision to focus on the last 5 months is also based on the institutional context: in this organization managers are asked to send the resignation letter three months before departing. We picked the starting point for our analysis two months prior to the official resignation – on month 5 – based on the assumption that the closer managers get to the final decision of quitting, the higher the likelihood to exhibit divergent communication behaviors.

In this study, we explore possible cues in the managers' communication behavior that indicate a change in the relationship "managers-organization" and possibly a fracture in the psychological contract. Following a method similar to the embeddedness approach to turnover (Mitchell, Holtom, Lee, Sablynski, & Erez, 2001) we used new social network metrics such as betweenness centrality oscillation, average response time, nudges and emotionality metrics (Allen, Gloor, Fronzetti Colladon, Woerner, & Raz, 2016; Gloor, Almozlino, Inbar, & Provost, 2014) to identify changes in the communication behaviors of managers who are close to quit their job.

Our study is embedded into a long history of examining the construct of turnover in terms of relationships (Feeley, 2000; Feeley & Barnett, 1997; Labianca & Brass, 2006; Mossholder, Settoon, & Henagan, 2005; Moynihan & Pandey, 2008; Soltis, Agneessens, Sasovova, & Labianca, 2013). While most of the previous studies have used the intention to leave as dependent variable, we correlate the actual number of managers leaving their job with measures of centrality, responsiveness to e-mail, language complexity and emotionality of the messages.

There is a lack of research examining the individual behavior that could lead to managerial turnover. While there are numerous empirical studies on the determinants and consequences of managerial turnover most of these studies focus on the role of environmental factors, firm profitability and strategic change (Brickley, 2003). Given the high costs associated with managerial turnover, such as the loss in firm-specific human capital and the costs of hiring a new manager (Sliwka, 2007), our method provides human resource departments with an effective tool to complement their incentive system and retention initiatives.

First, we review the existing literature on the determinants of managerial turnover, starting with the traditional attitude models and then focusing on the relational perspectives on turnover. Second, we describe our research design and the social network metrics used in our research: closeness, betweenness and degree centrality, oscillations in betweenness centrality, number of *nudges* sent and received, communication activity and average response time. Third, we discuss our hypotheses and report our empirical findings trying to identify managers who are likely to leave based on their communication patterns and managers who choose to stay. Finally, we discuss some practical implications, as well as limitations and opportunities to replicate and extend this study.

## 2. Traditional Determinants of Turnover

This section gives an overview of the literature on voluntary turnover and demonstrates the contribution of our approach, which looks at changes in the communication behaviors of managers before they leave the company. This overview of the literature on the main variables affecting turnover will also help provide empirical evidence to our selection of control variables. The variables most frequently reported as affecting turnover are usually falling into three categories: environmental/economic, organizational and individual (Selden & Moynihan, 2000). It has been shown that economic conditions might trigger voluntary turnover decisions, since employees are more likely to quit if they are confident that they will find easily another job (Cohen, 2003). Shih, Jiang, Klein and Wang (2011) found that increasing job autonomy can significantly reduce turnover, especially for jobs with a higher learning demand. In their meta-analyses of the main predictors of turnover, Griffeth, Hom, and Gaertner (2000) found that job satisfaction, organizational commitment and job involvement are the attitudinal variables most frequently investigated. Job satisfaction – which can be strongly influenced by job characteristics, even more than by personal motivation (Chen, 2008) – has been found to be the most reliable predictor of turnover: when employees express low job satisfaction, they are more likely to leave (Brawley & Pury, 2016; Cohen, 2003).

Several empirical studies have focused on the individual differences that could lead to a higher propensity to leave. It has been extensively demonstrated that the length of time in a position is negatively correlated with turnover (Cohen, 2003; Trevor, 2001). Two other demographic variables, race and gender, were usually considered major predictors of turnover, given the assumption that women and minorities would be more prone to quit. However, other researchers found that race and gender had scant predictive value on turnover when associated with other relevant variables (Lyness & Judiesch, 2001).

Some of the off-the-job factors that could possibly predict turnover include ample job opportunities and perceptions of the job market (Hom & Kinicki, 2001), family attachments (Lee & Maurer, 1999) or unpredictable events, also called shocks representing positive, negative or neutral events such as unsolicited job offers, changes in marital state, transfers, and firm mergers (Lee, Mitchell, Holtom, McDaniel, & Hill, 1999, p. 451).

Various reviews reported that attitudinal variables explain only about 4 to 5 percent of the variance in turnover (Griffeth et al., 2000; Mitchell et al., 2001). Although the traditional attitude approach to turnover has shown significant results, other significant factors should be included (Maertz & Campion, 1998). Some researchers suggested that turnover might be predicted looking at how well employees "fit" within the larger organizational culture (Mitchell et al., 2001). Villanova and colleagues (1994) predicted that a poor *employee-organization fit* was a good predictor of turnover, while O'Really, Chatman and Caldwell (1991) found that employees who did not fit within the culture quit their job faster than others, but only after 20 months of tenure.

What seems to be missing in traditional theory and research on voluntary turnover is the understanding that employees' decisions are based on the social relations they form within and outside their work environment. In the following section we explore more recent attempts to break away from the traditional categories of predictor variables, specifically job attitudes and ease of movement.

## 2.1. Relational Perspectives on Turnover

Researchers have been increasingly interested in examining turnover not exclusively on the basis of individual, organizational or environmental/economic factors. In the past fifteen years empirical results have been presented to account for the role of employee's social relationships in predicting voluntary turnover. A relational perspective on turnover has been attracting attention based on the assumption that social capital may increase job satisfaction and ultimately reduce turnover (Dess & Shaw, 2001). In their influential study, Krackhardt and Porter (1986) investigated communication ties between employees at a fast-food restaurant. The authors found that turnover was based on clusters of employees who occupied similar structural positions and communicated with each other more intensively.

There seems to be strong empirical evidence suggesting that embeddedness and strong relational ties, reflected by high network centrality are able to reduce voluntary turnover. Our study is inspired by the work done by Mitchell et al. (2001), who introduced job embeddedness as a new organizational attachment construct that was negatively correlated with voluntary turnover. Job embeddedness included individuals' links to other people, teams and groups, besides their perception of fit within the organization and their perceived sacrifice in case of voluntary turnover. Similarly to Mitchell et al. (2001), Maertz and Griffeth (2004) found that links to people and groups were negatively related to turnover. In this paper we identify social network metrics that could help predict who is actually leaving a company, rather than who is reporting the intention to quit.

Mossholder et al.(2005) proposed a relational model to explain turnover based on four attributes of intra-organizational relations: network centrality, coworker support, a sense of obligation toward coworkers, and interpersonal citizenship behavior. Their main assumption is that good relations with other employees increase the chance that individuals will stay in the organization. Similarly, Labianca and Brass (2006) speculate that negative intra-organizational relationships may reduce employee performance and chance for promotion and eventually encourage turnover.

Empirical research conducted by Moynihan and Pandey (2008) found that strong social intra-organizational networks reduce turnover intention. The reason could be that people who perceive a high level of support from coworkers feel some sort of responsibility toward them and are less likely to express their intention to leave. Contrary to their assumptions, the authors found a weak correlation between external networks and intention to quit. This is probably based on the index of dummy variables that they used to operationalize "professional activity", which included metrics such as attendance at national and local meetings, and whether employees read professional journals advertising job opportunities. As the authors note, another proxy for external social networks could bring a different story since "*professional involvement is only one type of external social network that individuals may rely on to find out about job opportunities*" (Moynihan & Pandey, 2008, p. 219).

Feeley and Barnett (1997) proposed an Erosion Model (EM) to predict employees' turnover based on their network position. They found that those employees who were more centrally located in the communication network tended to remain at their job, while those located on the periphery left their position or became even disconnected. When Feeley (2000) replicated the study to test the Erosion Model, he found that employees with high degree centrality or number of links in the network were less likely to turnover. A recent study on turnover intention conducted by Soltis et al. (2013) explored both workflow and advice network and demonstrated how certain types of ties are beneficial for keeping employees from quitting. The authors also found that an excessive number of certain ties actually increases employees' intention to quit. For instance, when an employee is being contacted by many others for work related issues, the employee's turnover intentions rise significantly (Soltis et al., 2013). Similarly, Oldroyd and Morris (2012) demonstrated that having many connections with other employees creates more communication demands, which have been associated with reduced thriving, burnout, collaborative overload and may ultimately contribute to turnover. In a recent study, Porter, Woo, and Campion (2016) found that internal networking behaviors are associated with a reduced likelihood of voluntary turnover, and external networking behaviors are associated with an increased likelihood of voluntary turnover.

This recent stream of research seems to recognize that intra-organizational social networks are indeed important to predict the likelihood of employees to quit. The connections we make at work become the ties that bind us to an organization and mediate the negative effect of factors that frequently lead to voluntary turnover (Moynihan & Pandey, 2008). Despite the recent interest on studying turnover using a relational perspective, there is still a lack of empirical evidence on which specific social network metrics are more likely to predict turnover. Social Network Analysis provides a method to investigate the information structure of organization, despite its main focus on the "structural dimension" of social capital (Goodwin & Emirbayer, 1994). To make sure we also captured the content of those interactions, we analyzed the content reported in the subject lines.

## 3. Hypotheses and Measures

In this paper we adopt a relational perspective on turnover, by exploring the properties of communication networks generated by managers who chose to leave the company and by others who decided to stay. Similarly to previous work done by Feeley (2000), Moynihan and Pandey (2008) and Soltis et al. (2013), we used three different network centrality metrics, based on the assumption that employees with higher ties to others are likely more embedded and bound to the organization and less likely to quit (Hahn, Lee, & Lee, 2015; Krackhardt & Porter, 1986; Mitchell et al., 2001). Figure 1 illustrates the theoretical model used in the current study.

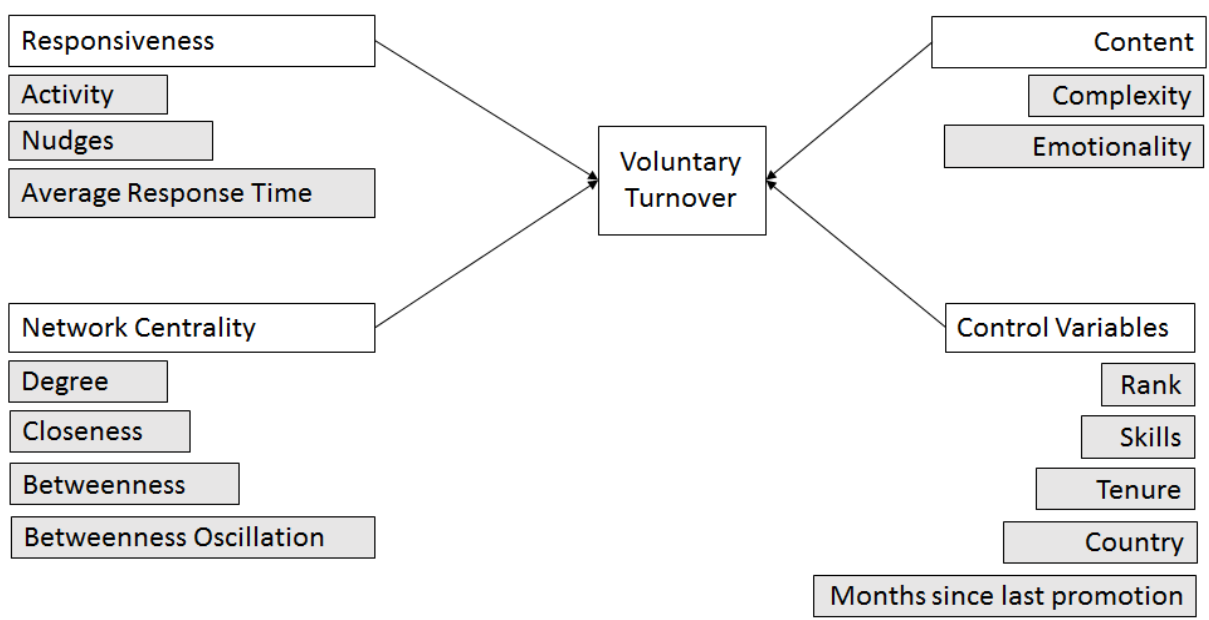

**Figure 1**. Theoretical model of voluntary turnover and communication network.

To operationalize centrality, we adopt three metrics which are well-known and commonly used in the social network analysis literature to identify dominant roles and prominence of actors (Kidane & Gloor, 2007; Wasserman & Faust, 1994). An e-mail network can be represented as an oriented graph composed of a set of $n$ nodes (e-mail accounts) – referred as G = {$g_1$, $g_2$, $g_3$ … $g_n$} – and of a set of $m$ oriented arcs (e-mails) connecting these nodes. The oriented graph can be represented by a sociomatrix $X$ made of $n$ rows and columns, where the element $x_{ij}$ positioned at the row $i$ and column $j$ is bigger than 0 if, and only if, there is an arc ($a_{ij}$) originating from the node $g_i$ and terminating at the node $g_j$. When the elements of X are bigger than zero, they represent the weight of the arcs in the graph ($x_{ij}$).

*Degree centrality* considers the number of arcs adjacent to a node, and in our network it represents the number of direct e-mail contacts of an employee. The higher the degree centrality, the higher the number of other people directly reached by that employee (Wasserman & Faust, 1994).

*Betweenness centrality* focuses on the capacity of a node to be an intermediary between any two other nodes. This measure is higher when an employee more frequently lies in the indirect communication patterns that interconnect other employees or people external to the company. A network is highly dependent on actors with high betweenness centrality, because of their position as intermediaries and brokers in the information flow (Borgatti, 2005). The betweenness centrality of the node $g_i$ is calculated counting the number of shortest paths linking all the generic pairs of nodes and dividing it by the number of paths which contain the node $g_i$ (Wasserman & Faust, 1994).

We also monitored *betweenness centrality oscillations* over time (Kidane & Gloor, 2007). An oscillation in betweenness centrality indicates that employees shift over time their active involvement in the communication flow, especially their role in transferring information from one person to another. Recent studies suggested that betweenness oscillation could be associated to higher levels of group creativity and be a predictor of success for joint projects between companies (Allen et al., 2016). A network with more oscillating leaders is usually more participative and less dominated by few individuals with a stable network position (Davis & Eisenhardt, 2011). We operationalize the measure of betweenness centrality oscillations counting the number of times a social actor changed his/her score of betweenness centrality (calculated weekly), reaching local maxima or minima, within the time interval of the study (Kidane & Gloor, 2007).

The other social network metric we used, *closeness centrality,* is based on the average length of the paths linking a node to others and reveals the capacity of a node to be reached, or to reach the others. The more central a node is, the shorter its communication paths. This measure can also be considered as a proxy of the speed at which a node can reach the others, without the need of relying on many peers to spread an idea or obtain an information. Closeness centrality is calculated as the inverse of distance of a node from all others in the network, considering the shortest paths that connect each pair of nodes (Wasserman & Faust, 1994).

In this paper, we posit that degree, closeness and betweenness centrality, as well as oscillation in betweenness centrality, are negatively correlated with voluntary turnover (**H1**). Managers with high centrality are usually more connected with others in the organization, have a higher membership stake and may be less prone to leave their job. Their greater involvement and more regular exchange with others make them more valuable members of the organization and sources of future assistance (Feeley, 2000; Sparrowe, Liden, Wayne, & Kraimer, 2001).

To identify a proxy for the level of engagement within the organization, we relied on network metrics developed specifically for e-mail networks. In particular, we looked at the communication *activity* via e-mail (Gloor et al., 2014), which indicates the number of e-mail messages sent by a person within a time interval, and *nudge*, which represents the number of pings (messages) a recipient receives before responding to an e-mail. We also further differentiate between ego nudges (i.e. number of pings before a recipient responds) and alter nudges (i.e. the number of pings before others respond). Our second hypothesis is based on the assumption that the more managers are involved in frequent interactions with others and are pinged more, the less likely they are to quit shortly after. This is aligned with the relational model proposed by Mossholder et al. (2005) who suggest that good relations among employees may help reduce the chance of turnover. An increased responsiveness to colleagues' e-mails might indicate the presence of stronger intra-organizational networks, a higher level of commitment to coworkers and therefore a likely reduction of turnover (Moynihan & Pandey, 2008). Therefore, we hypothesize that responsiveness - in the form of activity and nudges - is negatively correlated with voluntary turnover (**H2**).

We then use average response time (ART) to measure how much time it takes a person to reply to a particular e-mail (Gloor et al., 2014; Merten & Gloor, 2010). This metric is helpful to identify fast and slow communicators and possibly recognize patterns of behavior looking at periods of slower response. Merten and Gloor (2010) compared team satisfaction with average response time to e-mail and found that satisfied teams responded to e-mails somewhat faster. We expect managers to respond to e-mails at a slower rate when they are ready to leave a company, while their response time might be faster when they are actively working with peers and less distracted by outside job search activities (Moynihan & Pandey, 2008). Another reason to explain why employees – who are ready to quit - respond more slowly to e-mails is a possible burn out. As suggested by Soltis et al. (2013), when employees are being contacted by too many coworkers for work related issues, the employee's turnover intentions rise significantly, showing that some employees are being over-utilized.

Therefore, we postulate that voluntary turnover is positively correlated with average response time (**H3**). We further distinguish between ART-ego, which indicates the average time needed to answer an e-mail, and ART-alter, which represents the average time taken by others to respond to someone's e-mails. Both ART-ego and ART-alter are measured in hours.

Using the machine learning algorithms included in the social network and semantic analysis software Condor[1], we computed other two metrics: *complexity* and *emotionality* of the language used (Pang, Lee, & Vaithyanathan, 2002; Whitelaw, Garg, & Argamon, 2005). Using a multi-lingual classifier based on a machine learning method with data extracted from Twitter (Brönnimann, 2014) each e-mail in our archive was assigned with a *sentiment* value ranging from 0 to 1, where 0 denotes a negative sentiment, 1 a very positive sentiment and values around .5 a neutral one. Because sentiment is calculated as the average of the whole text, information can get lost. In order to capture the "*pathos*" transmitted by a message, we used another metric of sentiment analysis called "emotionality" (Brönnimann, 2014). Emotionality is measured as standard deviation of sentiment, i.e. the more fluctuations in positivity and negativity a message has, the more emotional it is. A second metric of sentiment analysis that we computed was the *complexity* of the language. Complexity denotes the deviation of word usage with the assumption that, the more we deviate from common, general language, the more complex is our language. Complexity is calculated as the likelihood distribution of words within a message, i.e. the probability of each word of a dictionary to appear in the text – using an algorithm based on the well-known term frequency/inverse document frequency information retrieval metric (Brönnimann, 2014). A message that uses more comparatively rare words has a higher complexity. Numerous studies support the idea that positive affectivity is associated with reduced intention to turnover, and that negative affectivity is associated with increased intention to turnover and actual turnover (Barsade & Gibson, 2007; Pelled & Xin, 1999; Thoresen, Kaplan, & Barsky, 2003). As illustrated by the Pennebaker (2013), our language can provide insights into our feelings and the application of computational linguistics represents an important tool to identify changes of emotional states. In particular, Pennebaker found that feelings of anxiety and sadness, which are typical during important life changing events such as quitting a job, tend to be expressed via the use of more inwardly focused words like "I, me, and my". Pennebaker (2013) also found that going through traumatic life events can lead to an increase use of I-words, a drop in big words, and an increase in the use of both positive and negative emotion words. Based on this, we would expect that a more emotional and complex language used by managers who quit, in the months prior to their departure, is connected to higher turnover (**H4**). By analyzing the content reported in the subject line, we were able to address all the components of the three dimensional model of social capital defined by Naphiet and Ghoshal (1998): the structural dimension, representing the connections among members, the relational dimension, which embraces cultural aspect and motivation of the relationships, and the cognitive dimension, which entails the content of the information flows.

---

[1] http://www.ickn.org/ckntools.html

Since this study involved a sample of respondents across different geographic areas, we decided to test the influence of the country where the managers were located. Other control variables included the manager's internal rank; months since last promotion, which could reduce job satisfaction and increase turnover; tenure within the company and skill (e.g. marketing, supply chain, Information Technology). We were not given data on gender and age of the participants; managers retiring naturally were not included in our sample.

### 3.1. Data Collection and Research Setting

The research setting was a large, global services organization operating in 25 countries, with key offices in the United States and more than 65,000 employees at the time of this study. We obtained access to e-mail data – with the possibility of fetching e-mail messages from the company servers – regarding 1566 managers who were employed at the beginning of data collection. The random sample was composed of 866 accounts. Each ego network was analyzed over 18 months, starting from October 2013. Out of the 866 managers involved in the study, we identified the 111 managers who left the company by the end of the observation period. Based on interviews with HR staff, the company did not experience any major organizational change during the observation period which could have led to an increase in turnover. Instead of asking managers their intention to quit, which is the typical surrogate variable for turnover (Lankau & Scandura, 2002; Mitchell et al., 2001; Moynihan & Pandey, 2008), we obtained the actual number of managers who left the company over a period of 18 months.

As a first step of the analysis, we compared the communication behavior of managers who worked throughout the whole period with the behavior of managers who resigned, in order to understand which variables could help forecasting a decision to leave the company. As a second step, we investigated the communication behavior of the 111 managers who left, comparing Months 1 to 13, with Months 14 to 16, right before their resignation.

## 4. Results

Our findings confirmed that managers who leave the company over the course of the eighteen months exhibit a different communication behavior when compared with their colleagues: their communication network is characterized by a lower closeness centrality, as well as a lower ego nudges and alter nudges activity.

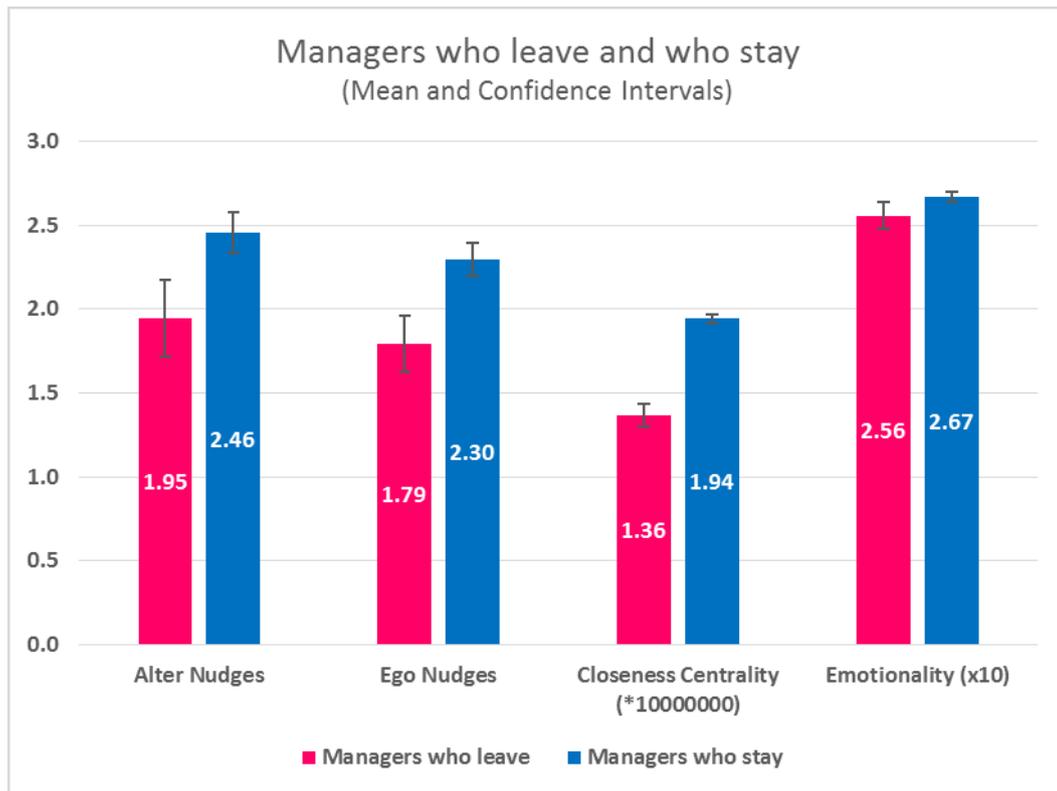

**Figure 2.** Managers who stay and who leave: independent sample t-test results ($p < .05$).

Turnover is therefore related to responsiveness, mainly in terms of nudges shared with colleagues: the more frequently managers interact with others and the more nudges they send and receive, the less likely they are to quit shortly after. Managers are also more likely to quit when their tenure is longer, though this effect is not as significant. Figure 2 illustrates significant independent sample t-tests ($p < .05$) that help distinguish between managers who stay and managers who quit their job.

Both the correlation results (Table 1) and the t-tests (Figure 2) suggest that closeness centrality and nudges (both alter and ego) acted as discriminating factors. Turnover was found to be negatively related to closeness ($r = -.477$, $p < .05$), to alter nudges ($r = -.104$, $p < .05$), ego-nudges ($r = -.124$, $p < .05$), and emotionality ($r = -.086$, $p < .01$). Managers who are getting ready to leave seem to be involved in e-mail exchanges that are more emotionally charged. The other variables in our models were not strongly correlated with voluntary turnover.

| | | Descriptive Statistics | | Correlations | | | | | | | | | | | | | | |
|---|---|---|---|---|---|---|---|---|---|---|---|---|---|---|---|---|---|---|
| | | M | SD | 1 | 2 | 3 | 4 | 5 | 6 | 7 | 8 | 9 | 10 | 11 | 12 | 13 | 14 | 15 |
| 1 | Managers who leave | 13% | | 1 | | | | | | | | | | | | | | |
| 2 | Rank | 1.25 | .43 | -.006 | 1 | | | | | | | | | | | | | |
| 3 | Tenure | 72.06 | 52.08 | .092** | -.049 | 1 | | | | | | | | | | | | |
| 4 | Months Since Last Promotion | 45.74 | 35.13 | -.012 | -.051 | .628** | 1 | | | | | | | | | | | |
| 5 | Activity | 60.04 | 293.64 | -.033 | .052 | .031 | -.019 | 1 | | | | | | | | | | |
| 6 | Alter ART | 22.52 | 17.29 | -.007 | -.005 | .011 | .032 | .019 | 1 | | | | | | | | | |
| 7 | Ego ART | 21.98 | 16.59 | -.040 | -.001 | -.014 | -.001 | .057 | .161** | 1 | | | | | | | | |
| 8 | Alter Nudges | 2.39 | 1.64 | -.104** | .058 | -.008 | -.005 | .096** | .231** | -.056 | 1 | | | | | | | |
| 9 | Ego Nudges | 2.23 | 1.35 | -.124** | .035 | -.129** | -.102** | .182** | -.009 | .165** | .069* | 1 | | | | | | |
| 10 | Betweenness | .00 | .01 | .010 | .101** | .100** | .016 | .814** | .040 | .039 | .021 | .134** | 1 | | | | | |
| 11 | Betweenness Oscillations | 4.40 | 2.39 | .047 | .188** | .169** | .053 | .152** | .070* | .063 | .057 | .060 | .158** | 1 | | | | |
| 12 | Degree | 7.83 | 12.25 | .008 | .111** | .123** | .016 | .833** | .073* | .067* | .081* | .177** | .941** | .352** | 1 | | | |
| 13 | Closeness | 1.87e-7 | 4.06e-8 | -.477** | .122** | -.072* | -.088** | .085* | .036 | .056 | .097** | .166** | .072* | .469** | .172** | 1 | | |
| 14 | Emotionality | .27 | .04 | -.086* | -.023 | .116** | .076* | .080* | .072* | .009 | -.025 | -.039 | .117** | .041 | .116** | .052 | 1 | |
| 15 | Complexity | 8.15 | .66 | -.060 | -.007 | .013 | .040 | .004 | .081* | .067* | .066 | .123** | -.014 | .058 | .013 | .129** | -.072* | 1 |

*$p < .05$; **$p < .01$.

**Table 1.** Distinguishing managers who decide to leave: Pearson's correlation matrix and descriptive statistics.

The logit regression confirmed these results as shown by the predictive models in Table 2. Some of the control variables, mainly tenure and months since last promotion, were somewhat relevant to predict turnover. We tested the differences accountable to the various ranks and jobs/functions managers had in the firm and the possible effect of the geographic area they were assigned. Our expectation was to observe a variance due to the country and to the level of unemployment in that country. A test by means of multilevel logistic regression found however that both specific job and country were not relevant determinants for the actual turnover. In the first models, we tested the significance of the predictors divided in blocks: control variables (Model 1), interaction variables (Model 2), centrality measures (Models 3-5) and language variables (Model 6). Due to collinearity problems, we could not include all the centrality measures in a single model (such a choice would determine a mean VIF of 6.72 and maximum VIF of 9.64). Lastly, we report two final models where we included the variables that in previous models were significant. In Model 8, the Mc Fadden's R-Squared is .323, and reductions in AIC and BIC scores are significant, demonstrating a good fit and proving the validity of our predictors.

| Variable | Model 1 | Model 2 | Model 3 | Model 4 | Model 5 | Model 6 | Model 7 | Model 8 |
|---|---|---|---|---|---|---|---|---|
| Rank | -.030 | | | | | | .361 | |
| Tenure | .008** | | | | | | .014** | .013** |
| Months Since Last Promotion | -.009* | | | | | | -.011** | -.010* |
| Ego Nudges | | -.484** | | | | | -.352** | -.346** |
| Alter Nudges | | -.330** | | | | | -.162* | -.164* |
| Alter ART | | .004 | | | | | | |
| Ego ART | | -.004 | | | | | | |
| Betweenness Oscillations | | | .058 | | | | | |
| Betweenness | | | .995 | | | | | |
| Closeness | | | | -37.605** | | | -39.778** | -39.602** |
| Degree | | | | | .002 | | | |
| Emotionality | | | | | | -6.375** | -4.231 | |
| Complexity | | | | | -.275 | | | |
| Constant | -2.148** | -.244 | 2.182** | 4.452** | -1.932** | 1.982 | 6.283** | 5.245** |
| Mc Fadden Adjusted R-Squared | .007 | .036 | -.006 | .283 | -.006 | .006 | .323 | .323 |
| AIC | 658.294 | 639.526 | 667.269 | 475.466 | 667.133 | 658.998 | 448.655 | 448.914 |
| BIC | 677.350 | 663.345 | 681.560 | 484.994 | 676.660 | 673.290 | 486.766 | 477.497 |
| N | 866 | 866 | 866 | 866 | 866 | 866 | 866 | 866 |

*$p < .05$; **$p < .01$.

**Table 2.** Logit models to identify the managers who decide to leave.

As an additional level of analysis, we studied how communication patterns changed within the sub-group of 111 managers who left the company by the end of the observation period. We analyzed their e-mail communications until six months before leaving, looking for signs of disengagement. We then compared these patterns with the communication behaviors of managers in the months 5 and 4, prior to leaving the company. By analyzing managers' e-mail communication during the months leading to their departure, we aimed to identify whether specific communication patterns could signal an emerging process of disengagement, or psychological detachment, measured with a resolution to leave (Burris et al., 2008). Figure 3 illustrates the significant differences in communication behaviors, tested by means of paired sample t-tests.

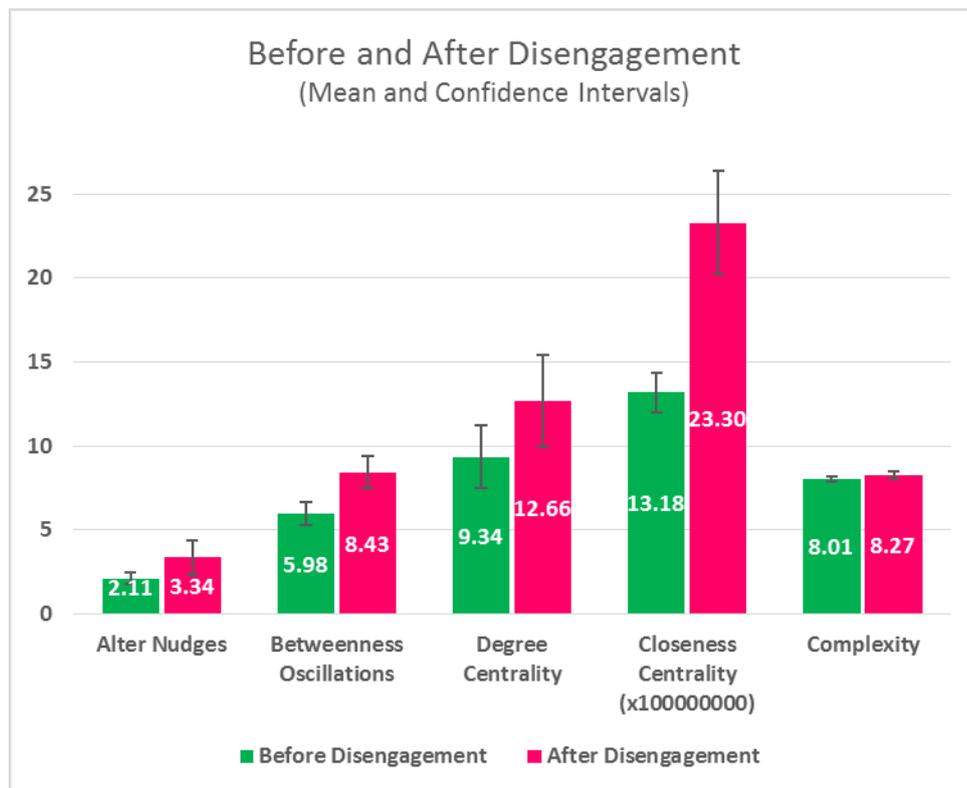

**Figure 3.** Change before and after disengagement: paired sample t-tests results ($p < .05$).

We found that the same variables that helped us differentiate managers who stay in their job from managers who quit can explain the difference in communication behaviors when a disengagement process emerges, and managers decide to leave. The other variables used in our theoretical model (Figure 1) were not statistically significant, thus we did not include them in Figure 3. The main variables that explain the changes in managers' communications are all network centrality metrics: degree centrality, closeness centrality and betweenness centrality oscillation. The results indicate that five months before leaving the company, managers pinged others more often (values of alter nudges are larger) meaning that others may not be as responsive as they used to. It seems that alters change their communication frequency towards managers who are going to leave soon. In order to completely understand the reason why other people interact differently with managers who are one step away from quitting we can only speculate that managers - who are dealing with some sort of disengagement at work - might also be changing their attitude towards the job and demonstrate this in their interaction with others. Consequently, other people might react to this change, even if they do not know what causes it. Following Lee & Mitchell's theory (1994), unhappy managers react to unexpected job-related events by starting mental deliberations about leaving, which could influence how they communicate with others and how others interact with them.

A possible reason for the increase in degree and closeness centrality is that managers - who finally decide to leave - might start reaching out and connect with people in their circle in order to explore job opportunities within the same company, in other departments or regions. Another explanation could be that they might connect again with their colleagues but now in a less engaged and committed way, since they are not under an obligation to perform; or, they may be wanting to leave in good terms with people who may be needed in the future. Lastly, it may just be that they have some separation anxiety and need reassurance and connection. The increase in betweenness centrality oscillation indicates that unhappy managers shift between highly central positions and more peripheral positions in the communication network right before they leave. This could be signaling their attempt to delegate tasks and responsibilities, as they are getting ready to quit their job.

## 5. Discussion

Our results suggest that managers who chose to leave the company are initially more actively involved in communicating with others, receive responses from others without pushing them too much (alter nudges is smaller) but are less central than managers who decide to stay. When we only focus on managers who are a couple of months away from leaving, we notice a turnaround in their communication behavior, as they are more central in the communication network than their colleagues, they rotate their role more frequently, and others need to be pushed (pinged) more to get a response.

The key contribution of this paper is the definition of a method that emphasizes the role of communication network metrics not traditionally associated with turnover. The sample of this study consisted of managers who voluntarily left their job at a large, global service company. This is a first, methodological contribution to the literature on turnover, since middle and top managers' decision to leave a company is investigated less frequently than employee turnover, mainly because of the small sample size that is usually involved. Managers have usually a larger membership stake and more to lose by leaving an organization. Overcoming the higher cost of leaving requires a strong inner drive to leave a process similar to a shock (Lee & Mitchell, 1994), to a psychological detachment (Burris et al., 2008) or disengagement (Kahn, 1990) that can prompt mental deliberations about leaving.

Our study offers a method to identify how online communication behaviors, specifically via e-mail, can be predictive of an emotional disengagement as a result of a psychological "shock". Shocks can be the effects of random events, unsolicited job offers, unexpected circumstances or luck into the quitting process. A shock can have a positive, neutral, or negative affect and differs from the concept of "unmet expectations" (Mowday, Porter, & Steers, 2013) which commonly involves newly hired employees during early employment periods. By monitoring over time metrics of social network analysis, we were able to recognize trends in managers' e-mail communication by simply monitoring online interactions. When managers make the decision to look for another job, they usually do not tell this to their supervisor until they have found a new job. Using a method similar to the one described in our study, managers in various organizations could reflect on their own online communication behavior and recognize the possible (positive or negative) impact on the organization.

While some network metrics were found to be more predictive than others in differentiating voluntary turnover, the results confirmed our main assumption that - once managers decide to leave - there is a process of disengagement that leads to a modification of their communication behavior.

Our results provided limited support for our first hypothesis (H1) since only closeness was significantly correlated to turnover. Oscillation in betweenness centrality was only significant to identify a change in behavior 5 months before leaving the job. The result on closeness seems to indicate that a greater involvement of managers and a regular communication with colleagues could reduce their intention to leave the job, which is aligned with previous studies (Feeley, 2000; Sparrowe et al., 2001). For example, the recent work of Porter et al. (2016) suggests that internal networking behaviors are associated with a reduced likelihood of voluntary turnover, while external networking behaviors are associated with an increased likelihood of voluntary turnover. This is an important result that supports previous empirical studies showing how a strong identification with the organization significantly predicted turnover intentions (Abrams, Ando, & Hinkle, 1998; Battistoni & Fronzetti Colladon, 2014). Our findings on centrality are also aligned with the results of previous studies (Feeley & Barnet, 1997; Sparrowe et al., 2001) which found that individuals who are close to others in the communication network have more direct and unmediated access to other organizational members. Those increased connections could translate into friendships and acquaintances ties which may serve to buffer the stress and tedium of everyday work (Feeley, 2000; Sias & Cahill, 1998). This study offers additional, empirical evidences to support a stream of literature that studies turnover using a relational perspective. The approach proposed in this paper uses social network metrics such as network centrality, activity and betweenness oscillation to learn more about typical communication patterns of managers who are close to break the psychological contract with the organization. Since the only presence of connections to others in the organization is not enough to prevent employees from quitting, we suggest using additional social network metrics to understand turnover. Previous studies have used mainly centrality measures to study the position of employees within their networks (Labianca & Brass, 2006; Mossholder et al., 2005; Soltis et al., 2013).

Of the other two metrics used to operationalize responsiveness – i.e. activity and nudges – only nudges was negatively correlated with voluntary turnover (H2). It seems that a simple metric like the number of e-mail messages sent by a manager before leaving (i.e. our activity measure) is not enough to predict turnover. Activity is an indicator that only measures how much an individual reaches out to others, while nudges – which was a good predictor of turnover – seems to be a better indicator of social dynamics by involving a respondent in the process. Since the decision to quit always involves a relationship between the individual and his/her context, including colleagues, family and potential new contacts, it makes sense that a more predictive value is offered by an intrinsically relational metric such as "nudge" rather than by an individual metric such as "activity".

Average response time does not seem to be a predictor of turnover, thus our third hypothesis is not supported (H3). Average response time has been associated to team satisfaction in previous studies using e-mail communications (Merten & Gloor, 2010). While we were expecting managers to respond to e-mails at a slower rate before leaving the company, due to a possible distraction for job-seeking activities or emotional disconnection, their response time did not change. It seems that the pattern showed by managers was one that favored the creation and maintenance of connections with colleagues, instead of progressive disconnection and separation.

Finally, we found partial support to our fourth hypothesis since language complexity seems to be the only good predictor of turnover, whereas emotionality is not. Emotionality, which measures the fluctuation in negative and positive terms used by managers in their e-mails, seems to be a good indicator to identify the managers who quit, but it does not seem to be able to predict when the disengagement starts to emerge. Managers who stay in their job tend to be less emotional in the language used in their e-mails, while their content emotionality increases over a long period of time if they eventually quit their job, not necessarily during the last few months. This is aligned with the research by Pennebaker (2013) showing that during life changing events there is an increase in the use of both positive and negative emotion words.

The emergence of a disengagement mechanism in the last five months on the job is strongly related to Lee and Mitchell's (1994) unfolding model of turnover, which discusses how "shocks" may prompt leaving behavior.

The empirical evidences collected in this study seem to suggest that voluntary turnover is associated with specific structural changes in the managers' social network position prior to their departure. It seems that managers who quit increase their "importance" in the communication network prior to their departure, by being closer to their colleagues. At the same time, they also need to "push" more in order to get a response from their colleagues, pinging others more often before receiving a response.

The theoretical contribution of this paper is to provide evidence that additional communication network metrics, such as betweenness centrality oscillation, closeness centrality, alter and ego nudges and language complexity should be studied to better understand managerial turnover. As this study demonstrates, managers change not only their communication frequency and their position in the communication network; they also tend to exchange e-mails that are more complex in the language used.

## 6. Managerial Implications

Overall, employee turnover, and in particular managerial withdrawal, are costly processes that can absorb as much as 17 percent of a company's income (Sagie, Birati, & Tziner, 2002). The detrimental effects on organizational performance have created a general interest in understanding the main determinants of voluntary turnover. Our study helps understand the emergence of a psychological disengagement by looking at how employees interact via e-mail. While this study was conducted at a large global service company, we believe that the results can be generalized to employees in other industries. Workers in different sectors - especially managers - are increasingly relying on digital communications to get their job done, thus we would not be surprised to obtain similar results in other knowledge-intensive industries (Shin & Choi, 2014). Managers and employees can benefit from looking at their own online communication patterns before they leave a company and reflect on their role in the communication networks. Studies like the one presented here could prompt a discussion on the implications of changing communication styles during critical situations, and can help individuals reflect on the ways their e-mail communication behavior could be interpreted by colleagues. E-mail is widely used as the main form of business communication as it helps increase efficiency, reduce geographic barriers and lower costs. At the same time, the reduced richness of e-mail communication can easily create misunderstandings if messages are not constructed properly, or if we do not respond promptly to certain receivers (e.g. important clients, supervisors) or do not connect with the necessary individuals. Our study demonstrates that managers change their online communication behaviors during critical stages of their employment, not connecting with others (reduce centrality) or becoming increasingly and suddenly connected, which can be misinterpreted by the receivers.

An important contribution of this paper is the proposal of a new method based on the actual number of managers who leave a company, instead of the most commonly used variable "intention to leave". Given the potential costs associated with managerial turnover, mainly represented by the loss in firm-specific human capital and the costs of hiring a new manager (Brickley, 2003; Sliwka, 2007), our method provides human resource departments with an effective tool to complement their incentive system and retention initiatives.

Similarly to the job embeddedness model (Mitchell et al., 2001), we illustrate the benefits of calculating social network metrics to observe individuals' intention to leave a company. In this paper we suggest the use of new metrics that could signal a disengagement process or inner termination that lead to managers to leave the company. Whereas Feeley (2000), Moynihan and Pandey (2008) and Soltis et al. (2013) used primarily network centrality metrics, our proposed method is based on additional metrics such as average response time, the number of times employees have to "nudge" the receiver before getting a response, or the oscillation over time in betweenness centrality. Another difference with previous studies that used a relational approach to turnover is that our method includes the calculation of metrics based on sentiment and content analysis, such as complexity of words used or emotionality of the message sent and received (Gloor & Giacomelli, 2014).

Using a similar approach to study actual turnover, human resource managers have the opportunity to rely on real time data regarding employees' communications. Using e-mail communication analysis, along with traditional methods to assess employees' satisfaction, human resource managers can offer the most appropriate organizational initiative such as mentoring programs, cross-staffing, or communities of practice that leverages the need for interdisciplinary efforts.

The application of the social network method described in this paper has the potential to help managers to better understand the nature of managerial turnover at their particular organization. This could inform the (re)design of turnover prevention strategies that fit with the organization's culture. It is not our intention to suggest a method to strictly monitor and control individuals' e-mail communication behavior: our method offers an opportunity to begin a self-reflection and plan proactive strategies that are tailored to the overall patterns observed in the past.

## 7. Limitations and Future Directions

It is important to replicate the study in a variety of other organizations, since relational variables may operate differently in various types of context. For example, researchers have found that in industries where turnover is high, like in fast food restaurants, developing strong ties with peers may aggravate turnover among employees with similar roles (Krackhardt & Porter, 1986).

When replicating this study, we also urge researchers to consider mapping the friendships and the advice network, which are important determinants of turnover intention (Bertelli, 2007). It is often the development of close friendships among employees and the loyalty to one another, rather than to the job or organization, that could discourage individuals from leaving a company (Sias & Cahill, 1998).

We also encourage future research to design a comprehensive study that takes into consideration traditional determinants of turnover along with relational indicators such as the ones proposed in this research.

Another limitation of this study is that we relied on data from a single sample which raises concerns about the generalizability of our findings. While this study is based on a single organization, the company we studied was a global organization operating in different industries and in 25 different countries. This makes the use of e-mail data even more appropriate for this organization, since managers communicate via e-mail more often than via phone or meetings (virtual or face-to-face). Due to institutional confidentiality issues, we based our analysis only on the subject lines of the e-mails exchanged during the 18 months. We recognize the intrinsic limitation in using only the subject of a message, instead of the whole e-mail message. Based on the publicly available Enron e-mail archive[2] we conducted an analysis similar to (Gloor & Niepel, 2006) and compared sentiment of subject line and message body of the 725 most active actors of the archive. The results show a significant correlation ($r = .25$) between sentiment of the subject line and sentiment of the message body.

Since our study focused exclusively on managers' communication behaviors, the findings cannot be generalized *tout court* to employees at all levels. Nevertheless, we would not be surprised to obtain similar results in other knowledge-intensive industries, where workers rely heavily on digital communications to get the job done.

---

[2] https://www.cs.cmu.edu/~./enron/

Finally, we want to recognize that monitoring organizational e-mails carries some ethical issues that should also be considered before conducting a similar research. Any organizational network analysis typically involves giving data or showing results to management. We do not encourage managers to target specific individuals whose communication behaviors show a different trend in centrality or responsiveness. Our intent was to educate managers and employees to recognize their own communication patterns before they leave a company and reflect on their position in the communication networks. Our concern is with protecting the individual respondents, and confidentiality and anonymity should be clearly embedded into any social network interventions (Borgatti & Molina, 2005).